\documentclass[aps]{revtex4}

\usepackage{amssymb}
\usepackage[tbtags]{amsmath}

\newcommand{\D}{\displaystyle}

\begin{document}
\title{Differential Transfer Relations of Physical Flux Density\\Between Time
Domains of the Flux Source and Observer}
\author{Ji Luo}
\email{luoji@mail.ihep.ac.cn}
\affiliation{Accelerator Center, Institute of High Energy Physics, Beijing, China}

\date{\today}

\begin{abstract} Differential transfer relations of flux density, general physics
quantities' and its corresponding energy's, between time domains of source and observer
are derived from conservative rule of various physical quantities and time function
between the two domains. In addition, integrated form of the relations are inferred and
examples of potential application are illustrated and discussed.
\end{abstract}
\maketitle

\section{Introduction}
Observing whole physical transportation process of a specified micro flux increment
from being emitted to detected during a pair of corresponding time intervals on time
domains of source and observer respectively, a fundamental fact can be realized that
both variations of physical quantity due to possible loss or energy exchange in the
process and transportational time for different flux signals within the micro flow
element can contribute detected flux density at observing location significantly.
Physical quantity, here divided into two main categories --- general physics quantity
such as phase of waves, mass, charge and particles' number and micro system's energy
carried by the micro flow element consisted of a kind of or a compound of general
physics quantities like a micro multi-charged particle system, contributes the
detectable flux density in two ways; for negative contribution, some quantity of the
micro flow which originated from source might be lost in the transportational process
and fails to reach the observer; for energy exchange related contribution to energy
flux density at observer, an amount of energy originated in external system joins or
adds to the micro system's energy for the system's energy gain till to reach the
observer, whereas it is converse for the system's energy release. Meanwhile a variation
of the flux signal's transportation time will result in a variation of time interval in
which the flux element is received by observer with respect to the time interval of the
flux element emission at source, so that it affects detecting flux density with time
factor. In addition, a variation of longitudinal length of the micro flux element along
its moving direction, as a particular physics quantity, exists in the process and
accompanied with their physical quantity's differential transfer relation
simultaneously. By combining conservative rule of physical quantity with a knowledge of
time function between time domains of signal source and observer~\cite{luoji:timefunction}, we will
set up the three differential transfer relations which are accompanied or combined for
any kind of physics flux between time domains of flux source and observer, and present
integral form of the differential transfer relations, their basic properties and their
potentially important application in a few cases.

\section{Derivation of the Differential Transfer Relations,
Basic Properties and Inference}

\subsection{Derivation of the Three Differential Transfer Relations}

For any micro continuous flow consisted of a kind of physical quantity from source to
observer, there are three essential differential transfer relations accompanied with
the transportation process simultaneously; they are the micro flux increment
themselves, corresponding longitudinal length of the micro flow and energy flux,
therefore the three transfer relations must share same time related characteristics or
time relations.

Specifically there is the transportation process as following:

A micro flux increment $\Delta Q(\dot {t})$ is released from instants $t - \Delta t$ to
$t$, then pass through the space between source and observer during transfer time $T(t
- \Delta t)$ and $T(t)$ for head and tail of the micro flow element respectively, and
finally are detected at observing location from instant $t' - \Delta t' = t - \Delta t
+ T(t - \Delta t)$ to instant $t' = t + T(t)$~\cite{luoji:timefunction}, here $\dot {t} \in [t,t +
T(t)]$. Consequently the differential transfer relations can be derived below.

\subsubsection{General Physics Flux Density's Differential Transfer
Relation}

From factual relation experienced by the micro flux increment $\Delta Q(\dot {t})$ in
transportation process, where $\Delta Q_l(\dot {t})$ denotes the loss amount of
physical quantity from $\Delta Q(\dot {t})$,

\begin{equation}
\label{eq201} \frac{d\Delta Q(\dot {t})}{d\dot {t}} = - \frac{d\Delta Q_l (\dot
{t})}{d\dot {t}} \qquad \dot {t} \in [t,t']
\end{equation}

\[
\int_t^{t'} {\frac{d\Delta Q(\dot {t})}{d\dot {t}}d\dot {t}} = - \int_t^{t'}
{\frac{d\Delta Q_l (\dot {t})}{d\dot {t}}d\dot {t}}
\]
or
\begin{equation}
\label{eq201b} \tag{\ref{eq201}$'$} \int_{\Delta Q(t)}^{\Delta Q(t')} {d\Delta Q(\dot
{t})} = - \int_0^{\Delta Q_l (t')} {d\Delta Q_l (\dot {t})}
\end{equation}
there
\begin{equation}
\label{eq202} \Delta Q(t') + \Delta Q_l (t') = \Delta Q(t)
\end{equation}
to multiply Eq.\ \eqref{eq202} with $\frac{1}{\Delta t}$ or $\frac{1}{\Delta t'}$, then
take a limit as $\Delta t$ and $\Delta t' \to 0$
%\begin{equation}
%\label{eq203}
%\begin{split}
%& \frac{dQ(t')}{dt'} + \frac{dQ_l (t')}{dt'} =
%\frac{dQ(t)}{dt}\frac{dt}{dt'}\\
%\text{or}\quad &J(t') + J_l (t') = J(t)\frac{dt}{dt'}
%\end{split}
%\end{equation}
\begin{eqnarray}
& \dfrac{dQ(t')}{dt'} + \dfrac{dQ_l (t')}{dt'} =
\dfrac{dQ(t)}{dt}\dfrac{dt}{dt'}\nonumber\\
\label{eq203} \text{or}\qquad &J(t') + J_l (t') = J(t)\dfrac{dt}{dt'}
\end{eqnarray}
here $J(t')> 0$, $J(t) > 0$. $J(t')$ and $J(t)$ are instantaneous flux or rate of
physical quantity's flow corresponding to instant $t'$at observing location and instant
$t$ at flux source respectively; $J_l (t')$ is rate of a total accumulated flux loss at
instant $t'$ at detecting position caused by the possible loss amount of physical
quantity during the transfer time $t' - t = T(t) = T(t')$; and

\begin{equation}
\label{eq203b}\tag{\ref{eq203}$'$} J_l (t') = \frac{dQ_l (t')}{dt'} = \int_t^{t'}
{\frac{d^2Q_l (\dot {t})}{d\dot {t}^2}} d\dot {t} = \int_0^{\frac{dQ_l (t')}{dt'}}
{d[\frac{dQ_l (\dot {t})}{d\dot {t}}]} \geqslant 0
\end{equation}

\subsubsection{Transfer Relation of Longitudinal Length of the
micro flow along moving direction}

As a particular physics flux, the length can be either compressed or elongated. So
except that the micro length $\Delta \ell(\dot {t})$ could be increased or lengthened
whereas the general micro increment $\Delta Q(\dot{t})$ can only be decreased in
general due to $\Delta Q_l(\dot{t})$, it has same transfer relations as general physics
flux has.

\begin{equation}
\label{eq204} \frac{d\Delta \ell(\dot {t})}{d\dot {t}} = - \frac{d\Delta \ell_c (\dot
{t})}{d\dot {t}} \qquad \dot {t} \in [t,t + T(t)]
\end{equation}
\[
 \int_t^{t'} {\frac{d\Delta \ell(\dot {t})}{d\dot {t}}} d\dot {t} = - \int_t^{t'}
{\frac{d\Delta \ell_c (\dot {t})}{d\dot {t}}d\dot{t}}
\]
or
\begin{equation}
\label{eq204b} \tag{\ref{eq204}$'$} \int_{\Delta \ell(t)}^{\Delta \ell(t')} {d\Delta
\ell(\dot {t})} = - \int_0^{\Delta \ell_c (t')} {d\Delta \ell_c (\dot {t})}
\end{equation}
\begin{equation}
\label{eq205} \Delta \ell(t') + \Delta \ell_c (t') = \Delta \ell(t)
\end{equation}
\[
\frac{d\ell(t')}{dt'} + \frac{d\ell_c (t')}{dt'} = \frac{d\ell(t)}{dt}\frac{dt}{dt'}
\]
\begin{equation}
\label{eq206} v_{io} (t') + v_c (t') = v_{sis} (t)\frac{dt}{dt'}
\end{equation}

Where $\Delta \ell_c (\dot {t})$ denotes the compressed length in transportation
process whereas $\Delta \ell_c (t') = \int_t^{t'} {\frac{d\Delta \ell_c(\dot
{t})}{d\dot {t}}d\dot {t}} $ denotes the total accumulated length of compression during
whole process $T = t' - t$; $v_c (t')$ is a rate of total accumulated compressed length
passed through observing point; when $\Delta \ell_c (\dot {t})$, $\Delta \ell_c (t')$
and $v_c (t')$ takes positive values they reflect compression related amount
respectively, whereas negative values express the opposites' of compression or
elongation related amount.

With a variation of the length, have a change of longitudinal density of the flux
quantity and relevant consequence been involved.

\begin{equation}
\label{eq206b}\tag{\ref{eq206}$'$}
\begin{split}
\frac{\rho (t')}{\rho (t)} &= \frac{\Delta Q(t')/\Delta \ell(t')}{\Delta Q(t)/\Delta
\ell(t)}= \frac{\Delta Q(t')\Delta \ell(t)}{\Delta Q(t)\Delta \ell(t')}\\
&=\frac{\Delta Q(t) - \Delta Q_l (t')}{\Delta Q(t)}\frac{v_{sis} (t)dt}{v_{io}
(t')dt'}\\
&=\left[ {1 - \frac{\Delta Q(t')}{\Delta Q(t)}} \right]\left[ {1 + \frac{\Delta \ell_c
(t')}{\Delta \ell(t')}} \right]
\end{split}
\end{equation}

This relates to interaction between the $\Delta Q(\dot {t})$ and external system and to
energy exchange in the transportation process.

\subsubsection{Energy Transfer and Exchange Relationvc\\ Respect to the
$\Delta Q(\dot {t})$}

From factual relations and defined restricted condition or terms in below

\begin{equation}
\label{eq207} \frac{d\Delta E(\dot {t})}{d\dot {t}} = \frac{d\Delta E_{exc} (\dot
{t})}{d\dot {t}} - \frac{d\Delta E_l (\dot {t})}{d\dot {t}} \qquad \dot {t} \in [t,t +
T(t)]
\end{equation}

\[
\int_t^{t'} {\frac{d\Delta E(\dot {t})}{d\dot {t}}d\dot {t}} = \int_t^{t'}
{\frac{d\Delta E_{exc} (\dot {t})}{d\dot {t}}d\dot{t}} - \int_t^{t'} {\frac{d\Delta E_l
(\dot {t})}{d\dot {t}}d\dot{t}}
\]
or
\begin{equation}
\label{eq207b}\tag{\ref{eq207}$'$}
\begin{split}
\int_{\Delta E(t)}^{\Delta E(t')} {d\Delta E(\dot {t})} &= \int_0^{\Delta E(t')}
{d\Delta E_{exc} (\dot {t})}\\
&\phantom{=}{}- \int_0^{\Delta E_l (t') = \Delta E(t) - \Delta E_{so} (t)} {d\Delta E_l
(\dot {t})}
\end{split}
\end{equation}
there
\begin{equation}\label{eq208}
\Delta E(t') - \Delta E_{exc} (t') + \Delta E_l (t') = \Delta E(t)
\end{equation}
and
\begin{equation}\label{eq209}
\begin{split}
&\frac{dE(t')}{dt'} - \frac{dE_{exc} (t')}{dt'} + \frac{dE_l
(t')}{dt'} = \frac{dE(t)}{dt}\frac{dt}{dt'}\\
\text{or}\qquad &\mathcal{J}(t') - \mathcal{J}_{exc} (t') + \mathcal{J}_l
(t')=\mathcal{J}(t)\frac{dt}{dt'}\\
\text{or}\qquad &{P}(t') - \mathcal{J}_{exc} (t') + \mathcal{J}_l
(t')={P}(t)\frac{dt}{dt'}
\end{split}
\end{equation}

Where $\Delta E(t)$ and $\Delta E(t')$ are the corresponding total energy carried by
the micro flux increment $\Delta Q(t)$ and $\Delta Q(t')$ respectively; $\Delta E_{so}
(t)$ is a fractional part of $\Delta E(t)$ and it is carried by $\Delta Q(\dot{t})$
from the flux source to detecting section as a contributed part of energy flux source
to observational energy flux at detecting position.

So exchange energy $\Delta E_{exc} (\dot {t})$ could have positive values for $\Delta
Q(\dot {t})$'s energy gain or negative values for $\Delta Q(\dot {t})$'s energy loss.

$\mathcal{J}(t')$,$\mathcal{J}(t)$are energy flux which correspond to
$J(t')=\frac{dQ(t')}{dt'}$ and $J(t)=\frac{dQ(t)}{dt}$ respectively; here $\mathcal
J(t')
> 0$, $\mathcal J(t)
> 0$, and
\begin{equation}\label{eq209a}\tag{\ref{eq209}$'$}
\mathcal J_l (t') = \frac{dE_l (t')}{dt'} = \int_t^{t'} {\frac{d^2E(\dot {t})}{d\dot
{t}^2}} d\dot {t}
\end{equation}

\begin{widetext}
\begin{equation}\label{eq209b}\tag{\ref{eq209}$''$}
\mathcal J_{exc} (t') = \frac{dE_{exc} (t')}{dt'} = \int_t^{t'} {\frac{d^2E_{exc} (\dot
{t})}{d\dot {t}^2}} d\dot {t}
\begin{cases}
>0&\quad\text{energy gain of }\Delta Q(t')\\
<0&\quad\text{energy loss of }\Delta Q(t')\\
=0&\quad\text{no energy exchange of }\Delta Q(t')\text{ with outside}
\end{cases}
\end{equation}
\end{widetext}

\subsection{Basic Properties --- Simultaneity, Independent
Conservation and Dependent on Rate of Signal Transfer Time}
\subsubsection{Simultaneity of the Accompanied Variation of $\Delta Q(\dot{t})$,
$\Delta\ell(\dot{t})$,$\Delta E(\dot{t})$, the Three Essential Transfer Relations and
Time Function Factor} Any change on the physical quantity in the micro increment,
longitudinal length, energy carried by $\Delta Q(t')$ and physical or differential
state inside of the micro flux element occurs simultaneously and is inner related.

\subsubsection{Independent Conservations of Physical Quantity and Energy}
Physical quantity and carried energy of a micro flux increment are independently
conserved in their transportation process (Ref. Eqs.\ \ref{eq201} and \ref{eq207}) and
their differential transfer relations between time domains of source and observers;

From equation (\ref{eq201b})
\[
\int_t^{t + T(t)} {\frac{d\Delta Q(\dot {t})}{d\dot {t}}d \dot{t}} = - \int_t^{t+T(t)}
{\frac{d\Delta Q_l (\dot {t})}{d\dot {t}}d \dot{t}}
\]
there its derivative for $t'$
\begin{equation}\label{eq210}
\begin{split}
&\frac{d\Delta Q(t')}{dt'} + \frac{d\Delta Q_l (t')}{dt'} = \frac{d\Delta
Q(t)}{dt}\left[ {1 - \frac{dT(t')}{dt'}} \right]
\\
\text{or}\qquad &\frac{d\Delta Q(t')}{dt'} + \frac{d\Delta Q_l (t')}{dt'} =
\frac{d\Delta Q(t)}{dt}\frac{dt}{dt'}
\end{split}
\end{equation}
then
\[
\int_{t_1 '}^{t_2 '} {\left[ {\frac{d\Delta Q(\dot {t})}{d\dot {t}} + \frac{d\Delta Q_l
(\dot {t})}{d\dot {t}}} \right]} d\dot {t} = \int_{t_1 }^{t_2 } {\frac{d\Delta Q(\dot
{t})}{dt}d\dot {t}}
\]
\[
\Delta Q(t_2 ') + \Delta Q_l (t_2 ') - \Delta Q(t_2 ) = \Delta Q(t_1 ') + \Delta Q_l
(t_1 ') - \Delta Q(t_1 )
\]
thus
\begin{equation}\label{eq211}
\Delta Q(t') + \Delta Q_l (t') - \Delta Q(t) \equiv 0
\end{equation}

Similarly, from equation (\ref{eq207b})
\[
\begin{split}
\int_{t' - T(t')}^{t'} {\frac{d\Delta E(\dot {t})}{d\dot {t}}d\dot {t}} &= \int_{t' -
T(t')}^{t'} {\frac{d\Delta E_{exc} (\dot {t})}{d\dot {t}}} d\dot {t}\\
&\phantom{=}{} - \int_{t' - T(t')}^{t'} {\frac{d\Delta E_l (\dot {t})}{d\dot {t}}}
d\dot {t}
\end{split}
\]
take its derivative for $t'$, then integration on corresponding $[t_1 ',t_2 ']$ and
$[t_1 ,t_2 ]$, there
\begin{equation}
\label{eq212} \Delta E(t') - \Delta E_{exe} (t') + \Delta E_l (t') - \Delta E(t) \equiv
0
\end{equation}

\subsubsection{Flux's Dependence on the Time Function Factor --- Rate of Signal Transfer Time}
Differential transfer relations of general physics flux and energy flux are time
functions' derivative dependent (Ref. Eqs.\ \ref{eq203} and \ref{eq207} ), here (Ref.\ %
~\cite{luoji:timefunction})\\ \begin{center} $t'(t) = t + T(t)$, $\dfrac{dt'}{dt} = 1 +
\dfrac{dT(t)}{dt}$; $T(t) = T(t')$\\\end{center} or \begin{center} $ t(t') = t' -
T(t')$,$\dfrac{dt}{dt'} = 1 - \dfrac{dT(t')}{dt'}$ \\\end{center} as the time function
is reversible and longitudinal length and relative velocity spread transfer relations
\begin{equation}
\label{eq213} \begin{split} \Delta \ell(t') - \Delta \ell(t) &=\left.\Delta \ell(\dot
{t})\right|_{t'} - \left.\Delta \ell(\dot {t})\right|_t \\
&= \int_t^{t'}\left[v_{is} (\dot{t},t - \Delta t) - v_{is} (\dot {t},t)\right]d\dot{t}
\end{split}
\end{equation}
there \begin{widetext}
\begin{equation}
\label{eq214} v_{io} (t')\frac{dt}{dt'} - v_{sis} (t) = \int_t^{t'} {\frac{\partial
v_{is} (\dot {t},r_{is} )}{\partial r_{is} }} v_{is} (\dot {t},t)\frac{dt_j }{dt}d\dot
{t}
\end{equation}
or
\begin{equation}\tag{\ref{eq214}$'$}
\label{eq214b} \frac{d\ell(t')}{d\ell(t)} = \frac{v_{io} (t')dt'}{v_{sis} (t)dt} = 1 +
\frac{1}{v_{sis} (t)}\int_t^{t'} {\frac{\partial v_{is} (\dot {t},r_{is} )}{\partial
r_{is} }} v_{is} (\dot {t},t)\frac{dt_j }{dt}dt
\end{equation}
with to multiply equations (\ref{eq213}) and (\ref{eq214}), then take limitation as
$\Delta t_j (\dot {t})$ and $\Delta t' \to 0$
\begin{equation}
\label{eq215}\left.\frac{d\Delta \ell(\dot {t})}{d\dot {t}}\right|_{t'} -\left.
\frac{d\Delta \ell(\dot {t})}{d\dot {t}}\right|_t = \int_t^{t'} \left[a_{is} (\dot
{t},t - \Delta t) - a_{is} (\dot {t},t)\right]d\dot {t}
\end{equation}
and

\begin{equation}
\label{eq216} \frac{\partial v_{is} (t',r_{is} )}{\partial r_{is} }v_{io}
(t')\frac{dt'}{dt} - \frac{\partial v_{is} (t,r_{is} )}{\partial r_{is} }v_{sis} (t) =
\int_t^{t'} {\frac{\partial a_{is} (\dot {t},r_{is} )}{\partial r_{is} }v_{is} (\dot
{t},t)\frac{dt_j }{dt}dt}
\end{equation}
or
\begin{equation}
\label{eq216b} \tag{\ref{eq216}$'$} \D{\frac{\D{\frac{d\Delta \ell(\dot {t})}{\Delta
td\dot {t}}\left| {_{t'} } \right.}}{\D{\frac{d\Delta \ell(\dot {t})}{\Delta td\dot
{t}}\left| {_t } \right.}}} = \D{\frac{\D{\frac{\partial v_{is} (t',r_{is} )}{\partial
r_{is} }}}{\D{\frac{\partial v_{is} (t,r_{is)} }{\partial r_{is} }}}}\D{\frac{v_{io}
(t')dt'}{v_{sis} (t)dt} }= 1 + \frac{1}{a_{sis}(t)+a_{is}( t,t)}\int_t^{t'}
{\frac{\partial a_{is} (\dot {t},r_{is} )}{\partial r_{is} }v_{is} (\dot
{t},t)\frac{dt_j (\dot {t})}{dt}d\dot {t}}
\end{equation}
\end{widetext}

Where $\Delta t_j (\dot {t})$ is time interval during which the $\Delta \ell(\dot {t})$
pass through a specific point, that has a relative distance $r_{is}$ with respect to
signal source; $v_{sis} (t)$ is initial relative velocity of head of the length with
respect to signal source, $v_{io} (t')$ is a relative velocity of the length's head
with respect to observing point; $a_{sis} (t)$ and $a_{is} (t')$ are corresponding
accelerations respectively.

\subsection{Major Inference From Eq.\ (\ref{eq203}) and (\ref{eq209})}

\subsubsection{General Physics Flux Transfer Function between
Two Time Domains}

\begin{equation}
\label{eq217} Q(t') + Q_l (t') = Q(t_0 ) + \int_{t_0 }^{t' - T(t')} {J(\dot {t})d\dot
{t}}
\end{equation}
or
\begin{equation}
\label{eq217b} \tag{\ref{eq217}$'$}\int_{t_0 '}^{t'} {[J(\dot {t}) + J_l (\dot
{t})]d\dot {t}} = \int_{t_0 }^t {J(\dot {t})d\dot {t}}
\end{equation}

\subsubsection{Energy Flux Transfer Function between The Time
Domains}

\begin{equation}
\label{eq218} \int_{t_0 '}^{t'} {[\mathcal J(\dot {t}) - \mathcal J_{exc} (\dot {t}) +
\mathcal J_l (\dot {t})]d\dot {t}} = \int_{t_0 }^t {\mathcal J(\dot {t})d\dot {t}}
\end{equation}
or
\begin{equation}
\label{eq218b}\tag{\ref{eq218}$'$} E(t') - E_{exc} (t') + E_l (t') = E(t_0 ) +
\int_{t_0 }^{t' - T(t')} {\mathcal J(\dot {t})d\dot {t}}
\end{equation}

Both of the transfer functions are ever increasing function.

\section{Illustration and Discussion}
\subsection{Examples of the Relations' Use}
\label{sec301}
\subsubsection{Phase Flux Density and Consequent Energy Flux Density}
For wave phase flux, $\frac{d\Delta \varphi _l (\dot {t})}{d\dot {t}} \equiv 0$, %
$\frac{d\varphi _l (t')}{dt'} \equiv 0$, from equation (\ref{eq203}) there
\[
 \omega'(t') = \omega(t)[1 - \frac{dT(t')}{dt'}]
\]
or
\begin{equation}
\label{eq301}\omega'(t')\left[ {1 + \frac{dT(t')}{dt'}} \right] = \omega(t)
\end{equation}
or
\begin{equation}
\label{eq302} \int_{t_0 '}^{t'} {\omega'(\dot {t})d\dot {t}} = \int_{t_0 }^t
{\omega(\dot {t})d\dot {t}}
\end{equation}
and from equation (\ref{eq217}), the phase transfer function
\begin{equation}
\label{eq302b} \tag{\ref{eq302}$'$}\varphi (t') = \varphi (t_0 ) + \int_{t_0 }^t
{\omega(\dot {t})d\dot {t}}
\end{equation}

$\omega'(t') = \frac{d\varphi (t')}{dt'}$ is observer's measuring angular frequency;
$\omega(t) = \frac{d\varphi (t)}{dt}$ is emitting angular velocity or frequency at
phase flux source for the same micro phase increment.

From equation (\ref{eq209}) and suppose $\mathcal J_{exc} (t') = \frac{dE_{exc}
(t')}{dt'} = 0$ in non-absorption media,
\begin{gather}
\label{eq303}
\begin{split}
\mathcal J(t') &=\mathcal J(t)\left[ {1 - \frac{dT(t')}{dt'}} \right] - \int_{t' -
T(t')}^{t'} {\frac{d\mathcal J_l(\dot {t})}{d\dot {t}}} d\dot
{t}\\
&=\mathcal J(t)\left[ {1 - \frac{dT(t')}{dt'}} \right] - T(t')\left.\frac{d\mathcal J_l
(\dot{t})}{d\dot {t}}\right|{_\xi }\\
&\hspace{10em} \xi \in [t' - T(t'),t']\\
\text{or}\qquad\qquad
%\label{eq303b}\tag{\ref{eq303}$'$}
%\begin{split}
P(t')&=P(t)\left[ {1 - \frac{dT(t')}{dt'}} \right] - T(t')\frac{d\mathcal J_l (\dot
{t})}{d\dot {t}}\left| {_\xi } \right.
\end{split}
\end{gather}

According to the electromagnetic wave energy flux relation, observer's measuring or
measured optical intensity not only has been determined by optical energy's spatial
diminishing factor related with $T(t)$, but also depend upon time function's factor
which is determined by a rate of signal transfer time. From equation (\ref{eq301}),
it's known that time function's effect must be considered in frequency-shift relevant
measures and data analysis.

\subsubsection{Charged Particles Flux and Energy Flux in Linac}
For charged particles flux in linear accelerator there exists the particles' number
relation of a micro increment
\begin{equation}
\label{eq304} \Delta N(t') + \Delta N_l (t') = \Delta N(t)\quad \text{and}\quad \Delta
N(t') \leqslant \Delta N(t)
\end{equation}
consequently
\begin{equation}
\label{eq304b}\tag{\ref{eq304}$'$} \Delta N(t')q + \Delta N_l (t')q = \Delta N(t)q
\quad \Delta N(t')q \leqslant \Delta N(t)q
\end{equation}

\begin{equation}
\label{eq304c} \tag{\ref{eq304}$''$}\Delta N(t')m_q + \Delta N_l (t')m_q = \Delta
N(t)m_q \quad \Delta N(t')m_q \leqslant \Delta N(t)m_q
\end{equation}

This means that the number of particles in the micro element will diminish as they are
accelerated.

However, according to the number's flux differential transfer relation

\begin{equation}
\label{eq305} \frac{dN(t')}{dt'} = \frac{dN(t)}{dt}\left[ {1 - \frac{dT(t')}{dt'}}
\right] - \frac{dN_l (t')}{dt'}
\end{equation}
at observing location instantaneous number flux can exceed the emitting number flux due
to time function related factor when $\frac{dT(t')}{dt'} < 0$.

Furthermore, as to rf accelerator $T(t)$ or $T(t')$ is periodic function with a
designated $\tau ^\ast $, $T(t + \tau ^\ast ) = T(t)$ and $T(t' + \tau ^\ast ) =
T(t')$; hence on different time domains along higher energy of the micro flux
increment, there will be ever increasing high frequency fractional factor in a Fourier
analysis on number flux, charge flux, mass flux and energy flux, although number of the
particles, and corresponding mass and charge of the specific micro increment will
decrease continuously in the transportation process.

According to equation (\ref{eq208}) the carried energy of the micro increment, $\Delta
E(\dot {t})$, can be progressively increased by positive $\Delta E_{exc} (\dot {t})$
values; then
\begin{equation}
\label{eq306} \Delta E(t') = \Delta E(t) + \Delta E_{exc} (t') - \Delta E_l (t')
\geqslant \Delta E(t)
\end{equation}

Compare equation (\ref{eq304c})
\[
\Delta M(t') = \Delta M(t) - \Delta M_l (t') \leqslant \Delta M(t)
\]
with equation (\ref{eq306}), we can make an inference that the energy gain $\Delta
E(t') - \Delta E(t)$ can not be stored as or converted into any mass of the micro
element for $\Delta M(t') - \Delta M(t) \leqslant 0$. Then where has been the energy
gain stored in the micro flux increment? We will find it stored as internal electrical
potential energy in the micro increment, and partially converted to the mass center's
kinetic energy of the micro increment~\cite{luoji:energy}.

\subsection{Potential Application}
\renewcommand{\labelenumi}{\Alph{enumi}.}
\begin{enumerate}
\item By means of phase transfer function (Ref.\ Eqs.\ %
\ref{eq301}, \ref{eq302}), the phase function on observer time domain at measuring
location of wave interference region can be transformed to phase source time domain on
which the angular frequency or velocity is known. Then phase difference on observer
time domain is expressed by term of time function and known or designated parameter of
phase source~\cite{luoji:phasedifference}.

\item Energy flux analysis on frequency-shift astronomical measurement~\cite{luoji:redshift}.
\item As an analytic approach one longitudinal properties, energy exchange and flux of
charged particles in acceleration~\cite{luoji:energy}.
\end{enumerate}

\section{Conclusion}
The differential transfer relations of general physics flux between time domains of
source and observer, accompanied with a longitudinal length's transfer relation, and
corresponding energy flux occur or exist simultaneously; and they reflect different
aspect of a specified flux increment's transportation process between source and
observer. Time function factor could significantly affect the values of corresponding
fluxes on two time domains and itself or rate of it is a necessary fraction of the
transfer relations. The differential transfer relations are set up on conversation of
various physics quantities independently and shared time transfer relation. Based on
the differential transfer relation of flux, its integral form --- physical quantity
transfer function between the time domains is inferred.

\end{document}